\documentclass[aps,prl,superscriptaddress,reprint]{revtex4-1}
\usepackage{amsmath}
\usepackage{amssymb}
\usepackage{graphicx}
\usepackage[colorlinks,urlcolor=blue]{hyperref}
\usepackage{url}
\usepackage{color,xcolor}
\usepackage{ulem}
\usepackage{float}
\usepackage{epstopdf}
\begin{document}

\title{First-principles study of the low-temperature charge density wave phase in the quasi-one-dimensional Weyl chiral compound (TaSe$_4$)$_2$I}
\author{Yang Zhang}
\email{Corresponding author: yzhang@utk.edu}
\author{Ling-Fang Lin}
\affiliation{Department of Physics and Astronomy, University of Tennessee, Knoxville, TN 37996, USA}
\affiliation{School of Physics, Southeast University, Nanjing 211189, China}
\author{Adriana Moreo}
\affiliation{Department of Physics and Astronomy, University of Tennessee, Knoxville, TN 37996, USA}
\affiliation{Materials Science and Technology Division, Oak Ridge National Laboratory, Oak Ridge, TN 37831, USA}
\author{Shuai Dong}
\affiliation{School of Physics, Southeast University, Nanjing 211189, China}
\author{Elbio Dagotto}
\email{Corresponding author: edagotto@utk.edu}
\affiliation{Department of Physics and Astronomy, University of Tennessee, Knoxville, TN 37996, USA}
\affiliation{Materials Science and Technology Division, Oak Ridge National Laboratory, Oak Ridge, TN 37831, USA}

\date{\today}

\begin{abstract}
Using {\it ab initio} density functional theory, we study the lattice phase transition of quasi-one-dimensional (TaSe$_4$)$_2$I.
In the undistorted state, the strongly anisotropic semimetal band structure presents two non-equivalent Weyl points.
In previous efforts, two possible Ta-tetramerization patterns were proposed to be associated with the low-temperature structure.
Our phonon calculations indicate that the orthorhombic $F222$ CDW-I phase is the most likely ground state for this quasi-one-dimensional system. In addition,
the monoclinic $C2$ CDW-II phase may also be stable according to the phonon dispersion spectrum. Since these two phases have very similar energies in our DFT calculations, both these Ta-tetramerization distortions likely compete or coexist at low temperatures. The semimetal to insulator transition is induced by a Fermi-surface-driven instability that supports the Peierls scenario, which affects the Weyl physics developed above $T_{\rm CDW}$. Furthermore, the spin-orbit coupling
generates Rashba-like band splittings in the insulating CDW phases.

\end{abstract}

\maketitle

\section{I. Introduction}

One-dimensional ($1D$) systems continue attracting considerable attention due to their rich physical properties and reduced
dimensional phase space. In particular, in low dimension systems the electron-electron, phonon-phonon, electron-phonon, and spin-phonon couplings are strongly
enhanced by the interactions between transition metal ions~\cite{DW,Monceau:ap}. Under some conditions,
free carriers may form charge density wave (CDW) or spin density wave (SDW) states due to the partial or complete condensation of excitations,
properties which are physically interesting and important for possible applications~\cite{Grioni:JPCM,Wang:prb,De Soto:prb}.

Several $1D$ bulk compounds have been widely studied. For example, with focus on the electronic correlation effects the Cu-oxide $1D$ ladders
were theoretically predicted and experimentally confirmed to have a spin gap and become superconducting~\cite{cu-ladder1,cu-ladder2,cu-ladder3,cu-ladder4}.
Recently, an analogous behavior was shown to develop in iron ladders BaFe$_2$$X_3$ ($X$=S or Se) that were reported to be superconducting at high
pressure~\cite{Takahashi:Nm,Zhang:prb17,Ying:prb17,Zhang:prb18,Zhang:prb19}. By considering phononic modes or spin-phonon instability, ferroelectric or multiferroelectric states
were predicted in some $1D$ systems~\cite{Lin:prm,Lin:prl,Cross:prb,Choi:prl,Zhang:arxiv,Dong:PRL14}.

(TaSe$_4$)$_2$I is a typical paradigmatic quasi-one-dimensional material that has been frequently studied for more than thirty
years \cite{Gressier:jsps,Gressier:Acb}. This system undergoes a Peierls phase transition at $263$~K, the so-called CDW transition, accompanied by an incommensurate structural distortion at low temperatures~\cite{Fujishita:ssc,Smaalen:jpcm,Favre-Nicolin:prl}. Above the critical temperature $T_{\rm CDW}=263$ K, the nearest neighbor (NN) Ta-Ta distances ($d_{\rm Ta-Ta}=3.206$ \AA ~at room temperature \cite{Gressier:Acb,Lorenzo:jpcm}) are identical along each chain. As expected, the CDW instability breaks the symmetry of the isometric chains leading to Ta-tetramerization modes \cite{Favre-Nicolin:prl,Lorenzo:jpcm}. Related experiments revealed that the Ta-tetramerization periodicity may be $c=4d_{\rm Ta-Ta}$ \cite{Smaalen:jpcm,Favre-Nicolin:prl,Favre-Nicolin:prl,Voit:sci}. In addition, this system is also considered to be a Weyl semimetal~\cite{Shi:arxiv,Li:arxiv}. More recently, possible axion physics was proposed for this special chiral Weyl material \cite{Gooth:nature,Schmeltzer:arxiv}. It also should be noted that Rashba-like band splittings may also occur in a chiral system \cite{Sante:prl}. Since the spin-orbit coupling (SOC) for Ta's $5d$ orbitals can be robust, such Rashba splitting is expected. All these developments in (TaSe$_4$)$_2$I provide
a unique and promising platform to display different interesting physical properties in a single material.

In previous works, two Ta-tetramerization patterns were proposed for the low-temperature structure, corresponding to the one-dimensional $B_1$ and $B_2$ representations \cite{Lorenzo:jpcm}. Since the Ta-tetramerization distortion is small, it can only be observed via weak reflections, and for this reason the lattice phase transition is still uncertain. For this reason, from the theoretical perspective {\it ab initio} phonon calculations for the superstructure could play an important role
in clarifying the real mechanism of lattice structural transition in this compound.

In the present publication, we perform first-principles density functional theory (DFT) calculations for (TaSe$_4$)$_2$I. First, our theoretical results indicate a strongly anisotropic electronic structure for the undistorted phase, corresponding to its quasi-one-dimensional geometry. Based on phononic dispersion calculations, we found that (TaSe$_4$)$_2$I
contains phonon softening instabilities in the $\Gamma_4$ mode, resulting into an orthorhombic distortion, which could correspond to the $B_1$ or $B_2$ representations. By considering the combined $B_1$+$B_2$ symmetry breaking, the system would decrease to the monoclinic $C2$ phase (No. $5$) with a pattern along chains involving
four different NN Ta-Ta distances. Our DFT calculations suggest that the CDW-I lattice phase is the most likely ground state for this system. Since the energy difference between the two CDW lattices, $B_1$ or $B_2$, can not be distinguished unambiguously, these two Ta-tetramerization phases are competitive at low temperatures. Our results also support that the semimetal-insulator transition is driven by the Fermi surface instability. Furthermore, for the heavy element Ta, moderate Rashba-like splitting bands are found in the low temperature CDW phases.

\begin{figure}
\centering
\includegraphics[width=0.48\textwidth]{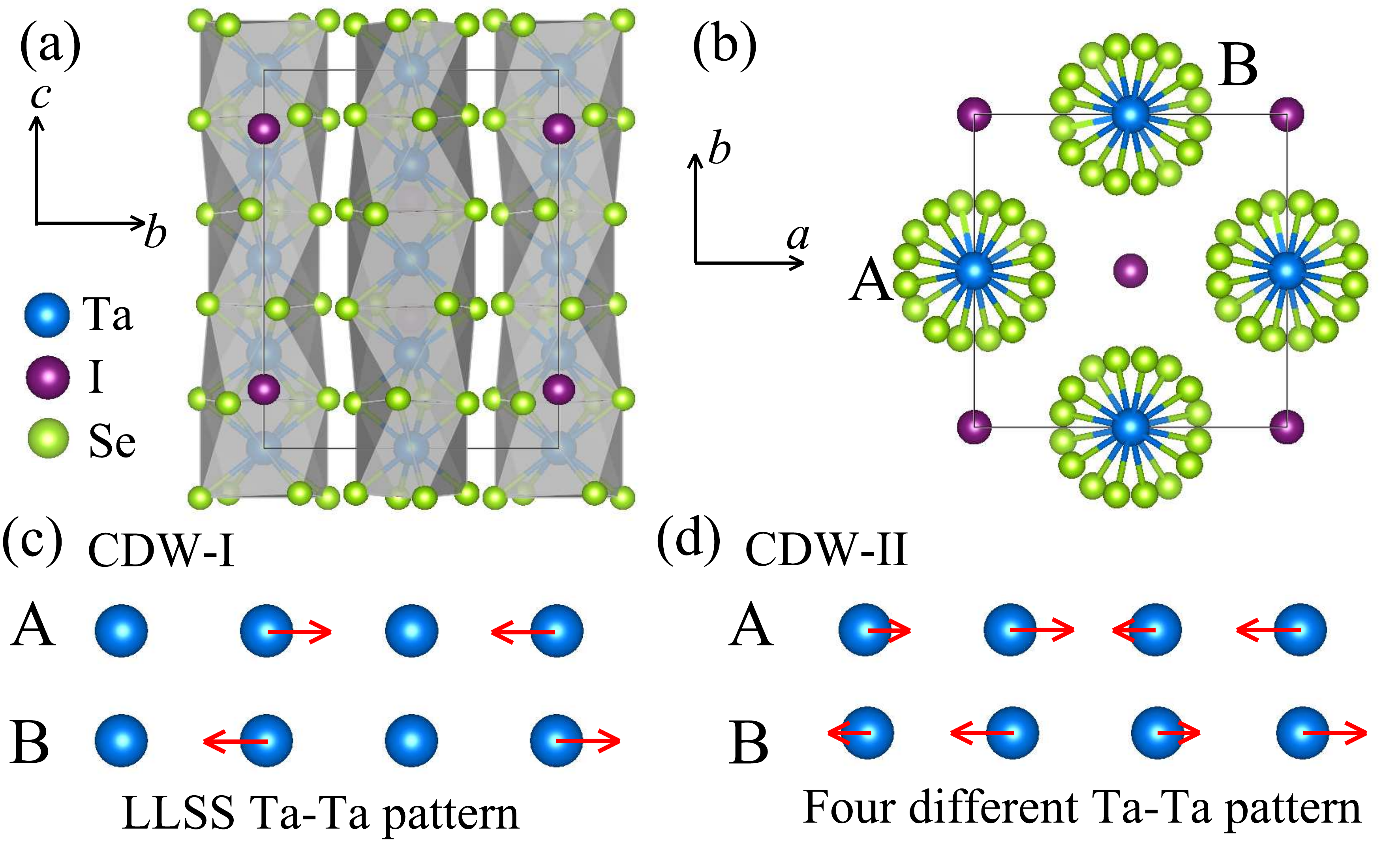}
\caption{Schematic crystal structure of the conventional cell of (TaSe$_4$)$_2$I with the convention: Blue = Ta; green = Se; purple = I. (a-b) The crystal structure is shown from the perspectives of the (a) $bc$ plane [($100$) direction] and (b) top [($001$) direction]. (c-d) The two different Ta-tetramerization patterns along the chains.}
\label{Fig1}
\end{figure}

\section{II. Methods and Model System}
In this publication, DFT calculations were performed using the projector augmented wave (PAW) pseudopotentials with the Perdew-Burke-Ernzerhof format
revised for solid exchange functional (PBEsol), as implemented in the Vienna {\it ab initio} Simulation Package (VASP)
code~\cite{Kresse:Prb,Kresse:Prb96,Blochl:Prb,Perdew:Prl}. The phonon spectra were calculated using the finite displacement approach and analyzed by the PHONONPY software \cite{Chaput:prb,Togo:sm}. Details are in the Supplemental Material (SM)~\cite{Supplemental}.

Under ambient conditions, (TaSe$_4$)$_2$I forms a quasi-one-dimensional body-centered tetragonal chiral crystal structure with the space group $I422$ (No. 97). The lattice parameters are $a=9.531$ \AA ~and $c=12.824$~\AA ~\cite{Gressier:Acb}. There are two adjacent TaSe$_4$ chains along the $c$-axis in the conventional cell, where iodine atoms are located between chains, as shown in Fig.~\ref{Fig1}(a). In each TaSe$_4$ chain, the Ta atoms are aligned equidistantly and the Se atoms form a ``screw'' arrangement along the $c$-axis [see Fig.~\ref{Fig1}(b)]. Previous experimental results suggested that the modulated structure of (TaSe$_4$)$_2$I can be split
into two parts~\cite{Lorenzo:jpcm} below the transition temperature $T_{\rm CDW}=263$ K.

Since the equal displacements of all atoms would not change the space-group  symmetry, we will focus on the modulation of the Ta-tetramerized atoms, which corresponds to the parallel component of the $k_F$ vector ($q_{\rm ||}^{\rm CDW}$) along the Ta-chain direction. By considering the space group symmetry $I422$ (No. $97$), the Ta atoms occupy two different Wyckoff sites ($4c$ and $4d$). Based on the Raman active modes analysis~\cite{Kroumova:pt}, the corresponding Raman active modes of the $4c$ and $4d$ sites have $B_2$ and $B_1$ symmetry, respectively. Since the CDW instability corresponds to the Ta-tetramerization modes \cite{Favre-Nicolin:prl,Lorenzo:jpcm}, a long-long-short-short (LLSS) Ta-tetramerization CDW-I pattern is expected by only considering the lattice dynamical $B_1$ or $B_2$ Ta tetramerization, as shown in Fig.~\ref{Fig1}(c) \cite{symmetrycontext}. If, instead, we consider the combined $B_1$+$B_2$ Ta tetramerization, there would be four different nearest-neighbor Ta-Ta distances in the chain [see Fig.~\ref{Fig1}(d)]. Based on the symmetry analysis, those two CDW phases corresponding to two Ta-tetramerization patterns are the most likely low-temperature structure. In addition, to better understand the Ta-tetramerization pattern along the chain direction, we calculated the phonon dispersion spectrum of a $1\times1\times4$ supecell for the parent phase based on the conventional cell. We only found that an unstable $\Gamma_4$ mode appears at $\Gamma$ (see Fig.S1) resulting in the LLSS Ta-tetramerization (CDW-I phase). Hence, we only considered those two states here. In addition, the AMPLIMODES software was employed to perform the group theoretical analysis~\cite{Orobengoa:jac,Perez-Mato:aca}, indicating that the CDW-I phase would induce the orthorhombic distortion from the undistorted tetragonal phase with phonon mode $\Gamma_4$. The same analysis for the CDW-II phase also indicates that a monoclinic distortion is obtained with phonon modes $\Gamma_5$ and $\Gamma_4$ corresponding to the undistorted tetragonal phase and the CDW-I orthorhombic phase, respectively.

\begin{figure}
\centering
\includegraphics[width=0.48\textwidth]{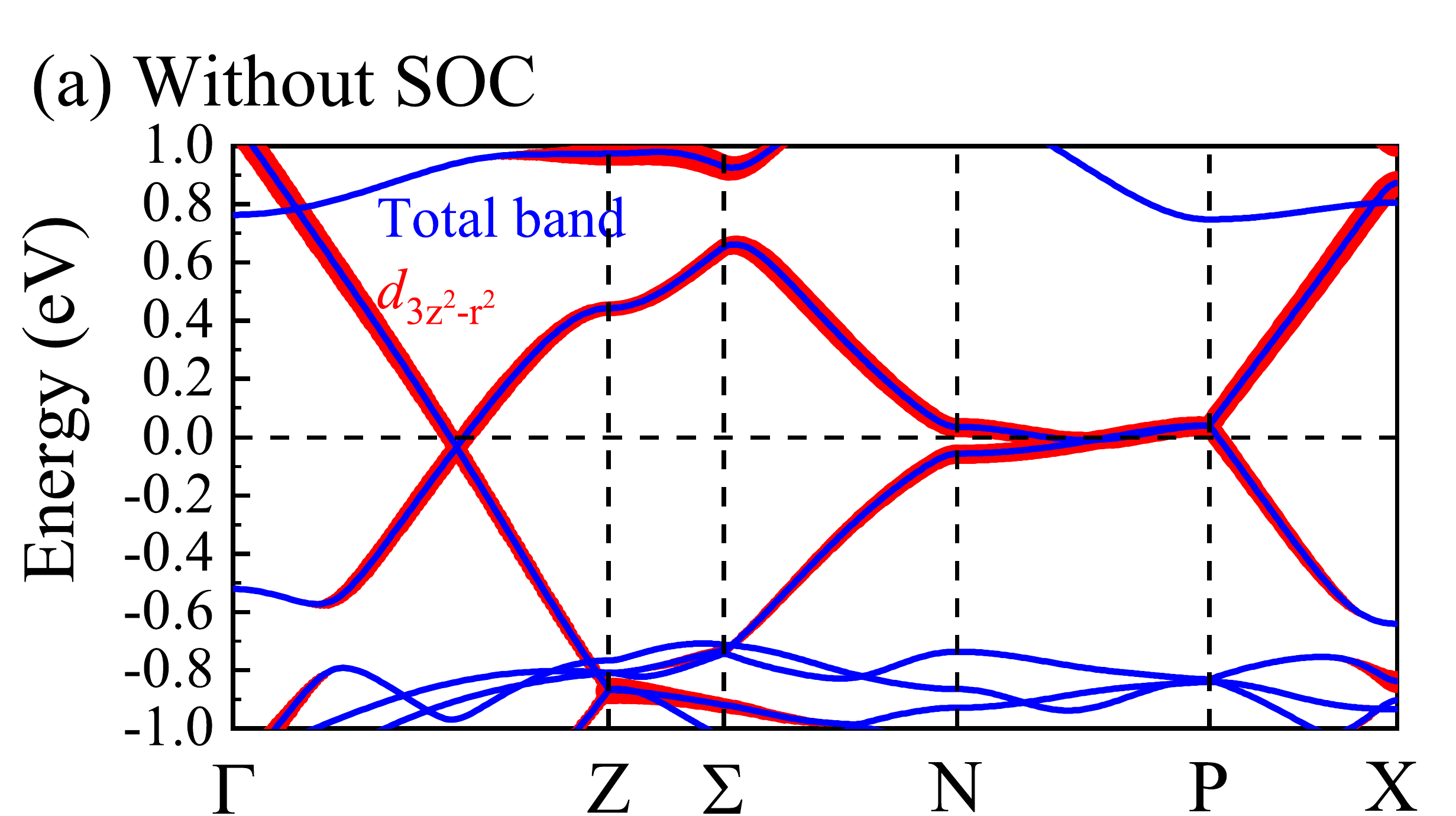}
\includegraphics[width=0.48\textwidth]{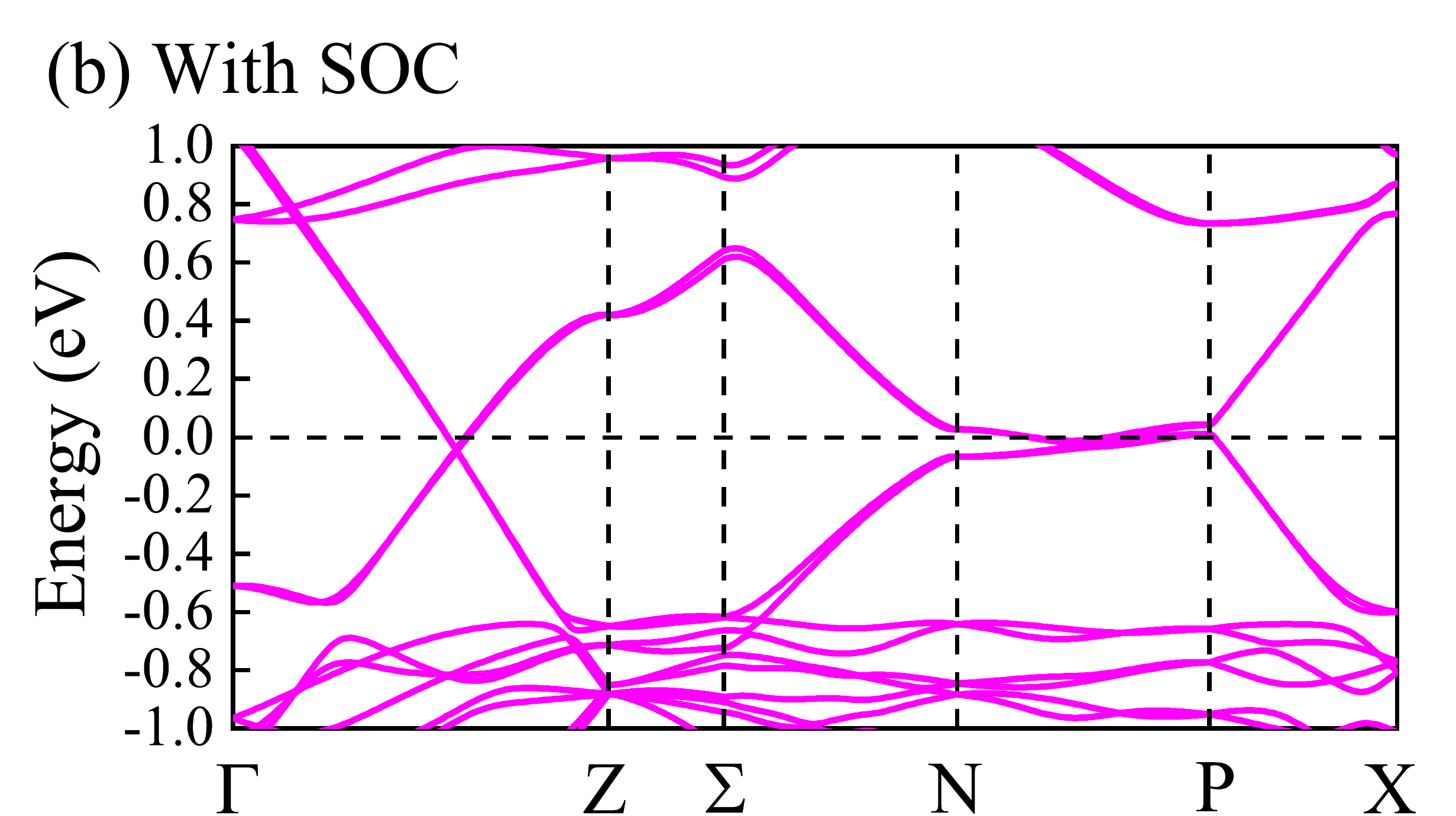}
\caption{Band structure of the undistorted (TaSe$_4$)$_2$I along a high symmetry path. (a) Without SOC. (b) With SOC.}
\label{Fig2}
\end{figure}

\section{III. Non-CDW phase}

Before discussing the CDW phase, we will consider the electronic structure corresponding to the non-CDW phase of (TaSe$_4$)$_2$I, which is displayed in Fig.~\ref{Fig2}. Here, we use the primitive cell instead of the conventional cell to calculate the electronic structure of (TaSe$_4$)$_2$I. The volume of the primitive cell is half of the conventional cell with the primitive lattice: $a_1$=($-a/2$, $a/2$, $c/2$), $a_2$=($a/2$, $a/2$, $c/2$), and $a_3$=($a/2$, $a/2$, $-c/2$), where $a$ and $c$ are the conventional cell-lattice constants.

First, we will present the band structure of the undistorted phase of (TaSe$_4$)$_2$I without SOC effect. This band structure clearly displays
strong anisotropic metallic behavior, as presented in Fig.~\ref{Fig2}(a). The band is more dispersive from $\Gamma$ to Z along the chains
than along other directions (i.e. N to P), which is compatible with the presence of quasi-one-dimensional chains along the $k_z$ axis. As shown in Fig.~\ref{Fig2}(a), the Fermi level is dominated by the Ta $d_{3z^2-r^2}$ orbital along the chains, in agreement with previous studies~\cite{Gressier:ic,Favre-Nicolin:prl}. The crossing point along the chain direction ($\Gamma$ to Z) is not at $\pi$/c but at the incommensurate $k_F$ where $2k_F$=$q_{\rm ||}^{\rm CDW}$.

Next, we introduce the SOC effect to the undistorted state and now the bands begin to split as displayed in Fig.~\ref{Fig2}(b), which is consistent with other results \cite{Li:arxiv,Shi:arxiv}. In principle, as suggested by the band structure, (TaSe$_4$)$_2$I should also display two interesting Weyl points with
one along the $\Gamma$ to Z path and another along the N to P path, although this aspect requires further detailed calculations and discussion beyond the scope of this publication. However, note that that the Weyl physics of this compound has already been studied in detail in recent publications ~\cite{Li:arxiv,Shi:arxiv}. For this reason, here we primarily focus on the phononic aspects.

\section{IV. CDW instability}

\begin{figure}
\centering
\includegraphics[width=0.48\textwidth]{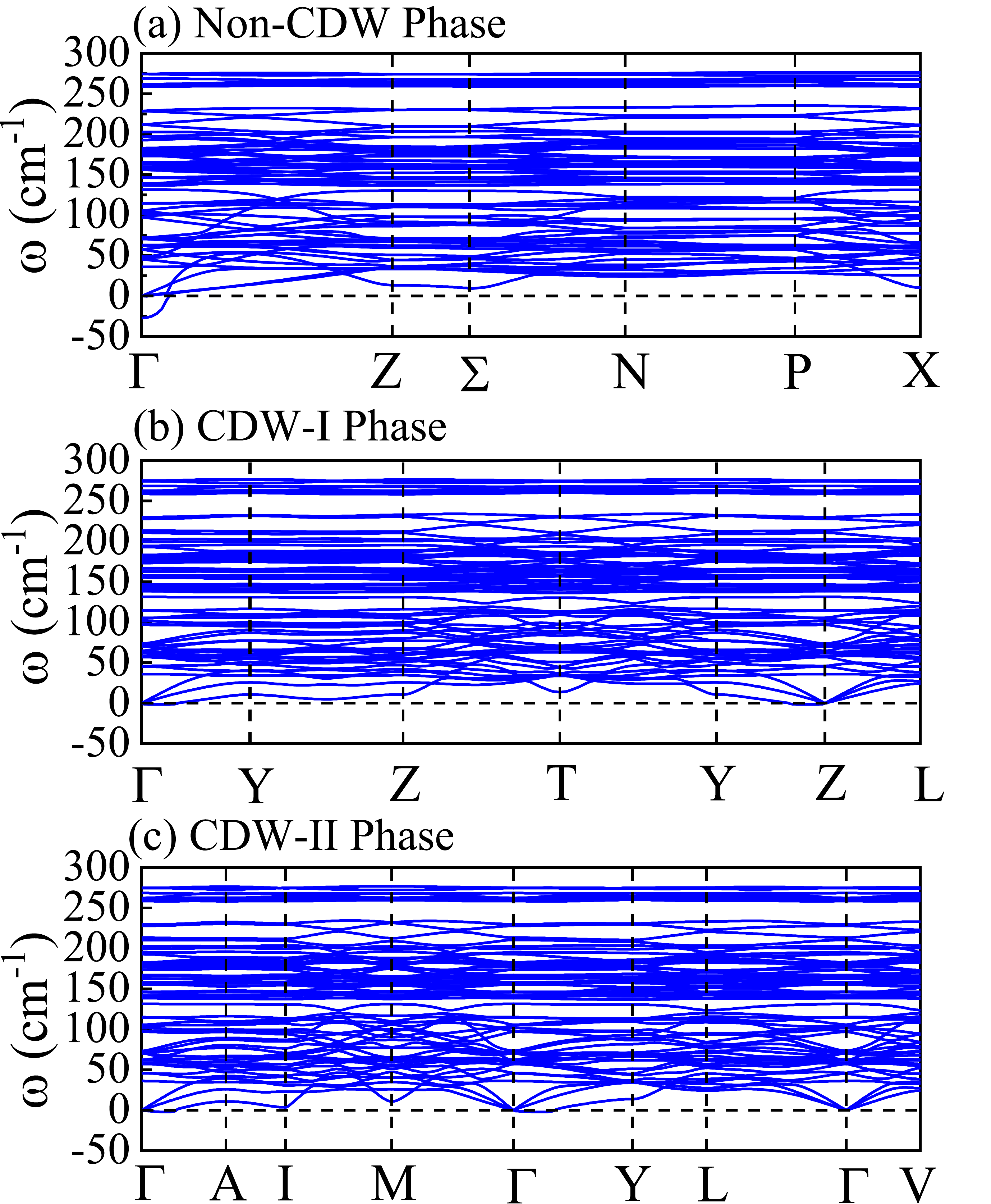}
\caption{Phonon spectrum of (TaSe$_4$)$_2$I for the undistorted (No. $97$), CDW-I (No. $22$), and CDW-II (No. $5$) phases.
Here, a $2\times2\times2$ supercell was selected from the original primitive cell in our calculations. The coordinates of the high symmetry points in the bulk Brillouin zone (BZ) are obtained using the Seek-path software \cite{Hinuma:cmc}.}
\label{Fig3}
\end{figure}

It is well known that the CDW instability is accompanied by an structural atomic rearrangement, related to the phonon mode instability. To understand the structural phase transition in (TaSe$_4$)$_2$I, we performed phononic dispersion calculations for both the non-CDW and CDW phases.  Figure~\ref{Fig3}(a) indicates the phonon dispersion spectrum that presents an imaginary frequency appearing at $\Gamma$ for the undistorted structure. The Ta-tetramerization arising from this mode corresponds to the CDW-I phase. According to group theory analysis using the AMPLIMODES software~\cite{Orobengoa:jac,Perez-Mato:aca}, this spontaneous distortion mode is a $\Gamma_4$ mode resulting in a twice larger periodicity ($4d_{\rm Ta-Ta}$) along the TaSe$_4$ chain~\cite{PHcontext}. In addition, the phonon softening mode also corresponds to the transverse component of the incommensurate CDW wavevector parallel to the chain direction ($\Gamma$ to Z), as in X-ray diffraction and ARPES data \cite{Lorenzo:jpcm,Tournier-Colletta:prl}.

We fully relaxed the crystal lattice from the original tetragonal $I422$ phase along these mode displacements, and then obtained an orthorhombic $F222$ state (No. $22$). Based on the relaxed orthorhombic structure, we found the LLSS Ta-tetramerization configuration with the difference between the long NN and short NN Ta-Ta distances being about $\sim 0.083$ \AA. Furthermore, we also calculated the phonon dispersion spectrum for the orthorhombic $F222$ phase (No. $22$), and in this case
no imaginary frequency mode was found, as shown in Fig.~\ref{Fig3}(b). By extracting the lowest-energy phonon mode of the orthorhombic $F222$ phase and applying to the $F222$ phase, the symmetry of the crystal structure further decreases to the monoclinic $C2$ phase (No. $5$) with a pattern of four different NN Ta-Ta distances along the chain, corresponding to the $B_1$+$B_2$ representations. The AMPLIMODES software~\cite{Orobengoa:jac,Perez-Mato:aca} indicates that this is the $\Gamma_4$ mode distortion for the orthorhombic $F222$ phase. After full lattice relaxations for the monoclinic $C2$ phase (No. $5$), we find that the difference of the two long or short Ta-Ta distances are $0.003$ \AA ~and $0.002$ \AA, respectively. Figure~\ref{Fig3}(c) indicates that the monoclinic CDW-II phase is also dynamically stable because no imaginary frequency mode is obtained in the phonon  dispersion calculation. Here, it should be noticed that the difference of energies of the orthorhombic and monoclinic cases is quite small ($\sim 0.1$ meV/Ta) which is beyond the accuracy of DFT. In this context, then those two phases cannot be distinguished by considering such small monoclinic distortion and almost negligible energy difference. Additional DFT results for those two phases can be found in the SM \cite{Supplemental}.

\begin{figure}
\centering
\includegraphics[width=0.48\textwidth]{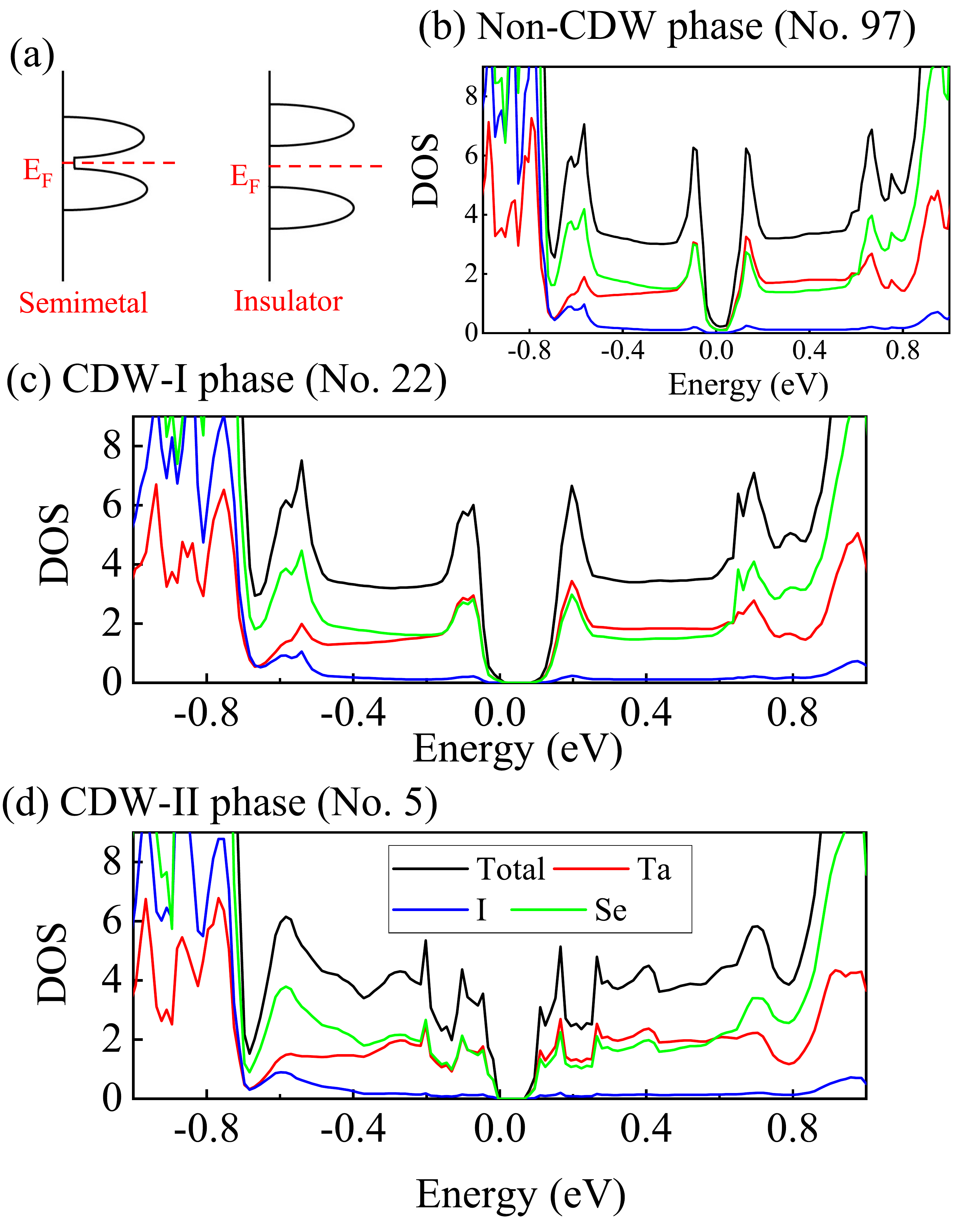}
\caption{(a) Schematic diagram of the  DOS semimetal to insulator transition. (b-d) DOS (without SOC) near the Fermi level for the undistorted parent
(non-CDW), orthorhombic (CDW-I), and monoclinic (CDW-II) phases, respectively. Black = Total; red = Ta; blue = I; green = Se. }
\label{Fig4}
\end{figure}

Next, let us focus on the semimetal to insulator transition. It is well known that there is a very small overlap between the bottom of the conduction band and the top of the valence band in the semimetallic material, resulting in a negligible density of states (DOS) without SOC at the Fermi level, as shown in Fig.~\ref{Fig4}(a).
The DOS of the undistorted phase (No. $97$) is here presented in Fig.~\ref{Fig4}(b) indicating a semimetal  with indeed quite a small density at the Fermi level. Furthermore, we also present in Figs.~\ref{Fig4}(c-d) the DOS for the distorted CDW phases showing that a gap opens in this case. By considering the SOC effect, the expected gap becomes approximately $0.2$~eV at the crossing point along $\Gamma$ to Z corresponding to the Brillouin path of undistorted phases. Here, the Ta-tetramerization plays a key role to understand the mechanism of the phase transition, which may be described by the Brillouin zone center Peierls instability. As shown in Figure~\ref{Fig2}, this system displays a quasi-one-dimensional band dispersion. Along the P-X path, parallel to the chains, the two Ta $d_{\rm 3z^2}$ bands are separated at the P point above the Fermi level with a small gap ($\sim40$ meV), resulting in a narrow hole pocket. Along the $\Gamma$-Z path, the two bands cross below the Fermi level, yielding a flat electron pocket. The pair is nested by the CDW wave parallel vector $q_{\rm ||}^{\rm CDW}$, consistent with the Peierls scenario. Hence, our results are in agreement with the physical picture of a Fermi-surface-driven instability resulting in the Peierls transition. Furthermore, strictly speaking the two Weyl points disappear in the low-temperature CDW phases due to the opening of a gap.  This suggests that Weyl physics could be altered below $T_{\rm CDW}$. Because the phase transition seems to be of first order, as suggested by
neutron scattering studies \cite{Lorenzo:jpcm}, the physics below and above $T_{\rm CDW}$ could be qualitatively different with regards to the Weyl features.

Then, we will briefly discuss the impact of doping effects in this system. Usually, for strong electron-phonon coupling CDW systems, a dilute isoelectronic doping should have little effect on the CDW transition temperature and CDW state because the location of the carries is determined by the electron-phonon coupling \cite{Schlenkerbook}. However, in contrast to other CDW systems, a surprising change of the modulation wave vector was reported in the Nb-doped (TaSe$_4$)$_2$I where a dramatic dependence of the satellite position was found at a small doping concentration $1\%$ ~\cite{Requardt:jpcm}. For non-isoelectronic doping, the doping effect can be described by the rigid-band picture where the conduction electron density is directly dependent on the dopant ions concentration.

\begin{figure}
\centering
\includegraphics[width=0.48\textwidth]{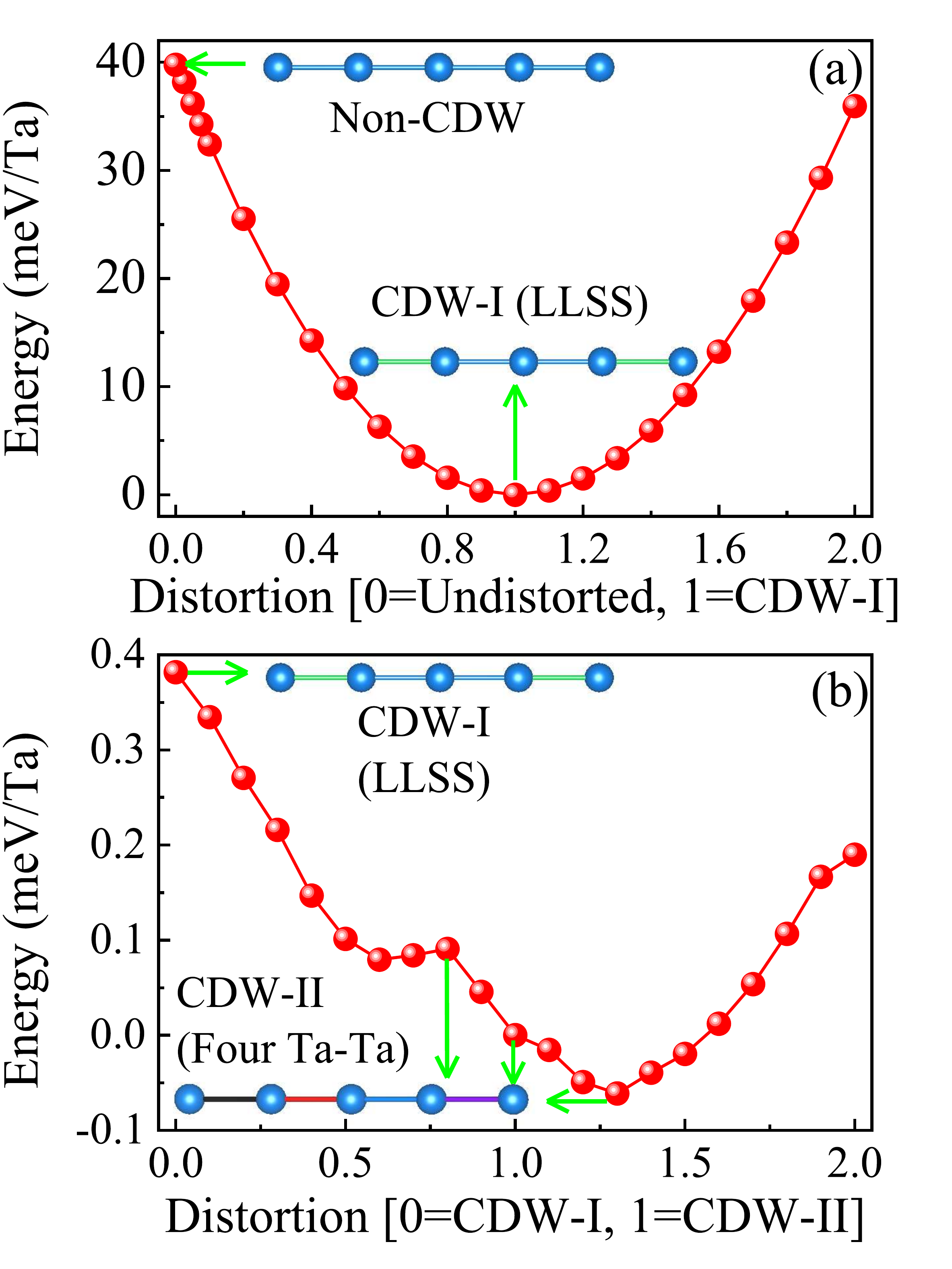}
\caption{(a) The energy well of the CDW-I and undistorted phases corresponding to the $\Gamma_4$ mode vs distortion.
(b) The energies well of the CDW-I and CDW-II phases.}
\label{Fig5}
\end{figure}

To understand better the lattice phase transition, we simulated the switching ``path'' from the undistorted phase (no Ta-tetramerization) to the CDW-I LLSS Ta-tetramerization phase by a simple linear interpolation of the $\Gamma_4$ mode, as shown in Fig.~\ref{Fig5}(a). The transition energy wall between the two phases was computed to be $\approx 40$ meV/Ta. However, note that we fixed the crystal constants to obtain the switching path and this would increase the energy barrier. In addition, we also compare the switching path between CDW-I and CDW-II phases and here the energy wall is much smaller $\approx 0.4$ meV/Ta [see Fig.~\ref{Fig5}(b)], which corresponds to a small monoclinic distortion.

\section{V. Additional discussion}

The point groups of (TaSe$_4$)$_2$I for the undistorted (space group No. $97$), CDW-I (space group No. $22$), and CDW-II (space group No. $5$) phases are $422$ ($D_4$), $222$ ($D_2$), and $2$ ($C_2$), respectively. In all these cases, optical isomerism is allowed in this quasi-one-dimensional system. For this reason,
optical activity effects are expected by considering their chiral point group. We thus suggest that further experimental and
theoretical works should focus on the optical isomerism of this material.

\begin{figure}
\centering
\includegraphics[width=0.48\textwidth]{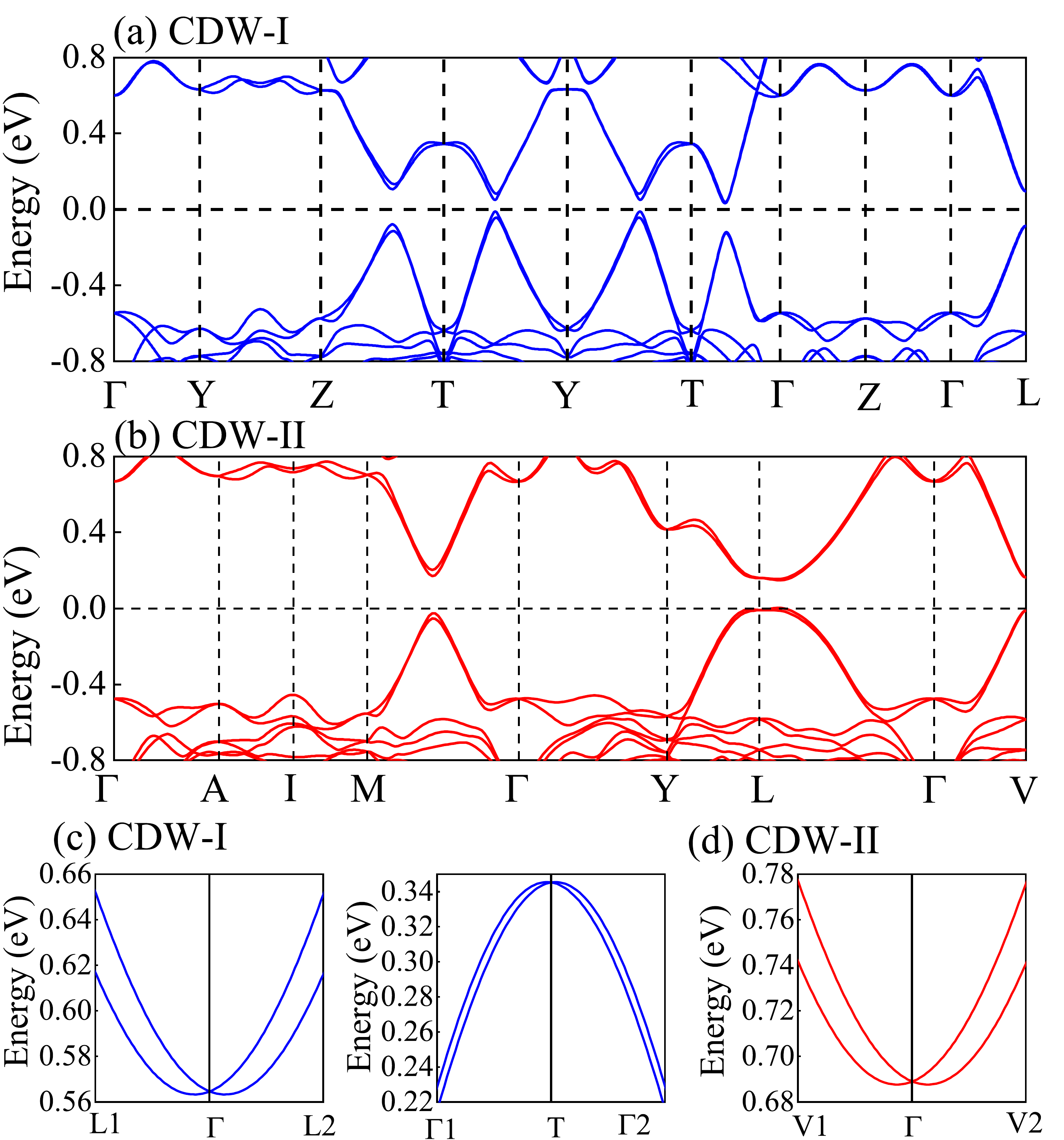}
\caption{Band structure of (TaSe$_4$)$_2$I with the effect of SOC shown along high symmetry paths. The coordinates of the high symmetry points in the bulk Brillouin zone (BZ) are obtained from the Seek-path software \cite{Hinuma:cmc}. (a) and (c) are for the orthorhombic $F222$ CWD-I phase. (b) and (d) are for
the monoclinic $C2$ CDW-II phase. }
\label{Fig6}
\end{figure}

Furthermore, for the heavy element Ta, the spin-orbit coupling for the $5d$ orbitals is expected to be large enough to induce a band splitting. Since the point groups of (TaSe$_4$)$_2$I for both the non-CDW and the CDW phases are the non-centrosymmetric chiral group, then a Rashba-like band splitting is also expected in this system. We calculated the band structures for the two CDW phases with SOC incorporated, with results in Fig.~\ref{Fig6}. There it is shown that the splitted bands resemble
Rashba splitting along the high symmetry directions, as shown in Figs.~\ref{Fig6}(a-b). Moreover, we also show the bands of CDW-I phases near the $\Gamma$ and T points that indicate a clear Rashba-like band splitting, as in Fig.~\ref{Fig6}(c). Similar Rashba-like bands splittings, see Fig.~\ref{Fig6}(d), indicate that these effects can also occur in the CDW-II phase. This is also reasonable since the monoclinic distortion is too small to completely alter the dominant physical properties. In fact, Rashba-like bands also develop along the $\Gamma$ to Z and N to P paths in the non-CDW parent phase. These issues deserve further studies in the future. It also should be noted that Shi et al. ~\cite{Shi:arxiv} found that the CDW couples the momentum-separated Weyl points with opposite chiral charge, providing a venue for the examination of topological insulating phases.

\section{VI. Conclusions}

In summary, here the quasi-one-dimensional compound (TaSe$_4$)$_2$I has been systematically studied using first-principles calculations. A strongly anisotropic metallic band structure was observed in the undistorted Weyl phase (non-CDW state), in agreement with previous experimental and theoretical works. In addition, using the group symmetry analysis and DFT calculations, two Ta-tetramerization CDW phases have been found here to become the stable low-temperature structure. The semimetal to insulator transition is induced by the Fermi-surface-driven instability, resulting in a Peierls transition that opens a gap ($\sim 0.2$ eV). In addition, based on the linear interpolation of the $\Gamma_4$ mode, the transition energy barrier is estimated to be $\approx 40$ meV, corresponding to a high transition temperature. Furthermore, for the heavy element Ta, the spin-orbit coupling for its $5d$ orbitals should be robust, causing a delicate but noticeable Rashba splitting of the conducting bands around the $\Gamma$ point in this polar system (chiral). Our results successfully reproduce the phase transition induced by the CDW instability and provide additional insight that may motivate further theoretical and experimental efforts. In future work we will explore the possibility of Rashba effects in these low-dimensional chiral materials.

\section{Acknowledgments}
E.D. and A.M. are supported by the U.S. Department of Energy (DOE), Office of Science, Basic Energy Sciences (BES), Materials Sciences and Engineering Division. S.D., Y.Z., and L.F.L. are supported by the National Natural Science Foundation of China (Grant Nos. 11834002 and 11674055). L.F.L. and Y.Z. are supported by the China Scholarship Council. Y.Z. is also supported by the Scientific Research Foundation of Graduate School of Southeast University (Grant No. YBPY1826).  All the calculations were carried out at the Advanced Computing Facility (ACF) of the University of Tennessee Knoxville.


\begin{references}
\bibitem{DW} G. Gr{\"u}ner, {\it Density Waves in Solids} (Perseus, Cambridge, MA, 2000).
\bibitem{Monceau:ap} P. Monceau, \href{https://doi.org/10.1080/00018732.2012.719674}{Adv. Phys. {\bf 61}, 325 (2012).}
\bibitem{Grioni:JPCM} M. Grioni, S. Pons and E. Frantzeskakis, \href{https://doi.org/10.1088/0953-8984/21/2/023201}{J. Phys.: Condens. Matter {\bf 21}, 023201 (2009).}
\bibitem{Wang:prb} Z. Wang and S.-C. Zhang, \href{https://link.aps.org/doi/10.1103/PhysRevB.87.161107}{Phys. Rev. B {\bf 87}, 161107(R) (2013).}
\bibitem{De Soto:prb} S. M. De Soto, C. P. Slichter, A. M. Kini, H. H. Wang, U. Geiser and J. M. Williams, \href{https://link.aps.org/doi/10.1103/PhysRevB.52.10364}{Phys. Rev. B {\bf 52}, 10364  (1995).}
\bibitem{cu-ladder1} E. Dagotto, J. Riera, and D. Scalapino, \href{https://doi.org/10.1103/PhysRevB.45.5744}{Phys. Rev. B {\bf 45}, 5744(R) (1992).}
\bibitem{cu-ladder2} E. Dagotto and T. M. Rice, \href{https://doi.org/10.1126/science.271.5249.618}{Science {\bf 271}, 618 (1996)}.
\bibitem{cu-ladder3} E. Dagotto, \href{https://doi.org/10.1088/0034-4885/62/11/202}{Rep. Prog. Phys. {\bf 62}, 1525 (1999).}
\bibitem{cu-ladder4} M. Uehara, T. Nagata, J. Akimitsu, H. Takahashi, N. Mori, and K. Kinoshita, \href{https://doi.org/10.1143/JPSJ.65.2764}{J. Phys. Soc. Jpn. {\bf 65}, 2764 (1996).}
\bibitem{Takahashi:Nm} H. Takahashi, A. Sugimoto, Y. Nambu, T. Yamauchi, Y. Hirata, T. Kawakami, M. Avdeev, K. Matsubayashi, F. Du, C. Kawashima, H. Soeda, S. Nakano, Y. Uwatoko, Y. Ueda, T. J. Sato and K. Ohgushi, \href{https://doi.org/10.1038/nmat4351}{Nat. Mater. \textbf{14}, 1008 (2015).}
\bibitem{Zhang:prb17} Y. Zhang, L. F. Lin, J. J. Zhang, E. Dagotto, and S. Dong, \href{https://doi.org/10.1103/PhysRevB.95.115154}{Phys. Rev. B \textbf{95}, 115154 (2017).}
\bibitem{Ying:prb17} J.-J. Ying, H. C. Lei, C. Petrovic, Y.-M. Xiao and V.-V. Struzhkin, \href{https://doi.org/10.1103/PhysRevB.95.241109}{Phys. Rev. B \textbf{95}, 241109(R) (2017).}
\bibitem{Zhang:prb18} Y. Zhang, L. F. Lin, J. J. Zhang, E. Dagotto, and S. Dong, \href{https://doi.org/10.1103/PhysRevB.97.045119}{Phys. Rev. B \textbf{97}, 045119 (2018).}
\bibitem{Zhang:prb19} Y. Zhang, L.F. Lin, A. Moreo, S. Dong, and E. Dagotto, \href{https://doi.org/10.1103/PhysRevB.100.184419}{Phys. Rev. B {\bf 100}, 184419 (2019).}
\bibitem{Lin:prm} L. F. Lin, Y. Zhang, A. Moreo, E. Dagotto, and S. Dong, \href{https://doi.org/10.1103/PhysRevMaterials.3.111401}{Phys. Rev. Mater. \textbf{3}, 111401(R) (2019).}
\bibitem{Lin:prl} L. F. Lin, Y. Zhang, A. Moreo, E. Dagotto, and S. Dong, \href{https://doi.org/10.1103/PhysRevLett.123.067601}{Phys. Rev. Lett. {\bf 123}, 067601 (2019).}
\bibitem{Cross:prb} M. C. Cross and D. S. Fisher, \href{https://doi.org/10.1103/PhysRevB.19.402}{Phys. Rev. B \textbf{19}, 402 (1979).}
\bibitem{Choi:prl} Y. J. Choi, H. T. Yi, S. Lee, Q. Huang, V. Kiryukhin, and S.-W. Cheong, \href{https://doi.org/10.1103/PhysRevLett.100.047601}{Phys. Rev. Lett. \textbf{100}, 047601 (2008).}
\bibitem{Zhang:arxiv} Y. Zhang, L. F. Lin, A. Moreo, S. Dong, and E. Dagotto \href{https://arxiv.org/abs/1912.07749}{arXiv preprint arXiv:1912.07749.}
\bibitem{Dong:PRL14} S. Dong, J. M. Liu, and E. Dagotto, \href{https://doi.org/10.1103/PhysRevLett.113.187204}{Phys. Rev. Lett. \textbf{113}, 187204 (2014).}
\bibitem{Gressier:jsps}P. Gressier, A. Meerschaut, L. Guemas, J. Rouxel and P. Monceau, \href{https://doi.org/10.1016/0022-4596(84)90327-X}{Journal of Solid State Chemistry \textbf{51}, 141 (1984).}
\bibitem{Gressier:Acb} P. Gressier and L. Guemas and A. Meerschaut, \href{https://doi.org/10.1107/S0567740882010176}{Acta Cryst. B \textbf{38}, 2877 (1982).}
\bibitem{Fujishita:ssc} H. Fujishita, M. Sato and S. Hoshino, \href{https://doi.org/10.1016/0038-1098(84)90576-3}{Solid State Commun. \textbf{49}, 313 (1984).}
\bibitem{Smaalen:jpcm} S. van Smaalen, E. J. Lam and J. L{\"u}decke, \href{https://doi.org/10.1088/0953-8984/13/44/308}{J. Phys. Condens. Matter \textbf{13}, 9923 (2001).}
\bibitem{Favre-Nicolin:prl} V. Favre-Nicolin, S. Bos, J. E. Lorenzo, J.-L. Hodeau, J.-F. Berar, P. Monceau, R. Currat, F. Levy and H. Berger, \href{https://link.aps.org/doi/10.1103/PhysRevLett.87.015502}{Phys. Rev. Lett. \textbf{87}, 015502 (2001).}
\bibitem{Lorenzo:jpcm} J. E. Lorenzo, R. Currat, P. Monceau, B. Hennion, H. Berger and F. Levy, \href{https://doi.org/10.1088/0953-8984/10/23/010}{J. Phys. Condens. Matter \textbf{10}, 5039 (1998).}
\bibitem{Voit:sci} J. Voit, L. Perfetti, F. Zwick, H. Berger, G. Margaritondo, G. Gr{\"u}ner, H. H{\"o}chst and M. Grioni, \href{https://science.sciencemag.org/content/290/5491/501}{Science \textbf{290}, 501 (2000).}
\bibitem{Tournier-Colletta:prl} C. Tournier-Colletta, L. Moreschini, G. Aut\`es, S. Moser, A. Crepaldi, H. Berger, A. L. Walter, K. S. Kim, A. Bostwick, P. Monceau, E. Rotenberg, O. V. Yazyev and M. Grioni, \href{https://link.aps.org/doi/10.1103/PhysRevLett.110.236401}{Phys. Rev. Lett. \textbf{110}, 236401 (2013).}
\bibitem{Gressier:ic} P. Gressier, M, H. Whangbo, A. Meerschaut and J. Rouxel, \href{https://doi.org/10.1021/ic00177a011}{Inorg. Chem. \textbf{23}, 1221 (1984).}
\bibitem{Shi:arxiv} W. Shi, B. J. WiedeR, H.L. Meyerheim, Y. Sun, Y. Zhang, Y. Li, L. Shen, Y. Qi, L. Yang, K. Jena, P. Werner, K. Koepernik, S. Parkin, Y. Chen, C. Felser, B. A. Bernevig and Z. Wang, \href{https://arxiv.org/abs/1909.04037}{arXiv preprint arXiv:1909.04037.}
\bibitem{Li:arxiv} X.-P. Li, K. Deng, B. Fu, Y. Li, D. Ma, J. Han, J. Zhou, S. Zhou and Y. Yao, \href{https://arxiv.org/abs/1909.12178}{arXiv preprint arXiv:1909.12178.}
\bibitem{Gooth:nature} J. Gooth, B. Bradlyn, S. Honnali, C. Schindler, N. Kumar, J. Noky, Y. Qi, C. Shekhar, Y. Sun, Z. Wang, B. A. Bernevig and C. Felser , \href{https://doi.org/10.1038/s41586-019-1630-4}{Nature \textbf{575}, 315 (2019).}
\bibitem{Schmeltzer:arxiv} D. Schmeltzer, \href{https://arxiv.org/abs/1910.13392}{arXiv preprint arXiv:1910.13392.}
\bibitem{Sante:prl} D. Di Sante, P. Barone, A. Stroppa, K. F. Garrity, D. Vanderbilt and S. Picozzi, \href{https://link.aps.org/doi/10.1103/PhysRevLett.117.076401}{Phys. Rev. Lett. \textbf{117}, 076401 (2016).}
\bibitem{Kresse:Prb} G. Kresse and J. Hafner, \href{https://doi.org/10.1103/PhysRevB.47.558}{Phys. Rev. B \textbf{47}, 558 (1993).}
\bibitem{Kresse:Prb96} G.~Kresse and J.~Furthm\"{u}ller, \href{https://doi.org/10.1103/PhysRevB.54.1169}{Phys. Rev. B \textbf{54}, 11169 (1996).}
\bibitem{Blochl:Prb} P. E. Bl\"{o}chl, \href{https://doi.org/10.1103/PhysRevB.50.17953}{Phys. Rev. B \textbf{50}, 17953 (1994).}
\bibitem{Perdew:Prl} J. P. Perdew and A. Ruzsinszky, and G. I. Csonka and O. A. Vydrov and G. E. Scuseria and L. A. Constantin and X. Zhou and K. Burke, \href{https://doi.org/10.1103/PhysRevLett.100.136406}{Phys. Rev. Lett. \textbf{100}, 136406 (2008).}
\bibitem{Chaput:prb} L. Chaput, A. Togo, I. Tanaka, and G. Hug, \href{https://doi.org/10.1103/PhysRevB.84.094302}{Phys. Rev. B \textbf{84}, 094302 (2011).}
\bibitem{Togo:sm} A. Togo, I. Tanaka, \href{https://doi.org/10.1016/j.scriptamat.2015.07.021}{Scr. Mater. \textbf{108}, 1 (2015).}
\bibitem{Supplemental} For more results, see Supplemental Material at \href{http://link.aps.org/supplemental/10.1103/PhysRevB.xx/xxxxxx}{http://link.aps.org/supplemental/10.1103/PhysRevB.xx/xxxxxx.}
\bibitem{Kroumova:pt} E. Kroumova, M. I. Aroyo, J. M. Perez-Mato, A. Kirov, C. Capillas, S. Ivantchev, and H. Wondratschek, \href{https://doi.org/10.1080/0141159031000076110}{Phase Transitions \textbf{76}, 155 (2003).}
\bibitem{symmetrycontext} The two Ta sites ($4c$ and $4d$) form a pattern Ta(1)-Ta(2)-Ta(1)-Ta(2) in the chain direction. In the $B_1$ representation, two Ta(1) atoms ($4c$ site) move towards the connected Ta(2) atoms in chain-A and move away from the connected Ta(2) atoms in chain-B, resulting in a LLSS Ta-tetramerization pattern. Correspondingly, in the $B_2$ representation, the Ta(2) atoms ($4d$ site) move towards the connected Ta(1) atoms in chain-A and move away from
the connected Ta(1) atoms in chain-B. Both the $B_1$ and $B_2$ representations would induce a simlar LLSS Ta-tetramerization CDW phase, as shown Fig.~\ref{Fig1}(c).
\bibitem{Orobengoa:jac} D. Orobengoa, C. Capillas, M. I. Aroyo, and J. M. Perez-Mato, \href{https://doi.org/10.1107/S0021889809028064}{J. Appl. Crystallogr. \textbf{42}, 820 (2009).}
\bibitem{Perez-Mato:aca} J. Perez-Mato, D. Orobengoa, and M. I. Aroyo, \href{https://doi.org/10.1107/S0108767310016247}{Acta Crystallogr. A \textbf{66}, 558 (2010).}
\bibitem{Hinuma:cmc} Y. Hinuma, G. Pizzi, Y. Kumagai, F. Oba and I. Tanaka, \href{https://doi.org/10.1016/j.commatsci.2016.10.015}{Comput. Mater. Sci. \textbf{113}, 221 (2016).}
\bibitem{PHcontext} Here, for completeness we also used $1\times1\times2$ and $1\times1\times4$ supercells based on the conventional cell, but none of our physical conclusions changed, as shown in the SM \cite{Supplemental}.
\bibitem{Schlenkerbook} C. Schlenker, Low-dimensional Electronic Properties of Molybdenium Bronzes and Oxides (Kluwer Academic Publishers, Dordrecht, 1989.)
\bibitem{Requardt:jpcm} H. Requardt, J. E. Lorenzo, R. Currat, P. Monceau, B. Hennion, H. Berger and F. Levy, \href{https://link.springer.com/book/10.1007/978-94-009-0447-7}{J. Phys. Condens. Matter \textbf{10}, 6505 (1998).}
\end{references}
\end{document}